\documentclass[a4paper,11pt]{article}
\usepackage{graphicx,amssymb,amsmath,amsfonts,epsfig}
\usepackage[usenames,dvipsnames]{color}
\usepackage[linktocpage,pagebackref]{hyperref}
\usepackage{subfigure}
\usepackage{xcolor}
\title{Distinguishing Signature of Kerr-MOG Black Hole and Superspinar via Lense-Thirring Precession}
\author{Parthapratim Pradhan\footnote{pppradhan77@gmail.com}\\ 
{\it Department of Physics}\\
{\it Hiralal Mazumdar Memorial College For Women}\\
{Dakshineswar, Kolkata-700035, India}}
\date{}

\begin{document}

\maketitle

\begin{abstract}
We examine the geometrical differences between the black hole~(BH) and naked singularity~(NS) 
or superspinar via Lense-Thirring~(LT) precession in  spinning modified-gravity~(MOG).  
For \emph{BH} case, we show that the LT precession frequency~($\Omega_{LT}$) along the pole  is proportional 
to the angular-momentum~($J$) parameter or spin parameter~($a$) and is inversely proportional to the cubic 
value of radial distance parameter, and also governed by Eq.~(\ref{1.1}).
Along the equatorial plane it is governed by Eq.~(\ref{1.2}).
While for \emph{superspinar}, we show that the LT precession frequency is inversely proportional to the cubic 
value of the spin parameter and it decreases with distance by MOG parameter as derived in Eq.~(\ref{1.3})
at the pole and in the limit $a>>r$ (where $\alpha$ is MOG parameter). For $\theta\neq\frac{\pi}{2}$ 
and in the superspinar limit, the spin frequency varies as $\Omega_{LT}\propto \frac{1}{a^3\cos^4\theta}$ 
and by Eq.~(\ref{ns1.2})

\end{abstract}


\section{Introduction:}
The fundamental difference between a black hole~(BH) and a naked singularity~(NS)~
\footnote{When no event horizons are formed during gravitational collapse then  it is called as NS 
or superspinar~\cite{Nakao18}. It is unstable compact object.} is that the former has a horizon 
structure while the latter doesn't have any horizon structure. Once more a BH is of two types: 
non-extremal black hole~(NXBH) and  
extremal black hole~(XBH). A NXBH is identified  by non-zero surface gravity~($\kappa \neq 0$) 
while XBH is identified by \emph{zero} surface gravity~($\kappa=0$). For NS the surface gravity 
is undefined because it has no horizon. So, how one can  differentiate between these three compact 
objects? This is an intringuing topic of research both in astrophysics and in general relativity. 
Therefore, in this Letter we want to find out that the distinction between these three compact 
objects by using LT precession frequency~\cite{lt,mtw,schiff,prl11}.  

For spinning Kerr BH in general relativity, the LT precession frequency varies as
spin parameter and is inversely proportional to the cubic value of radial distance both along the 
polar axis and the equatorial plane. While for Kerr NS or superspinar, the LT frequency varies as 
radial distance and inversely proportional to the cubic value of the spin parameter along the pole. 
This is the fundamental difference between BH and superspinar in the general relativity.  Now we will 
examine what happens this situation in case of other alternative gravity theory i.e. Kerr-MOG gravity?

Consequently we have considered a spinning compact object such as Kerr-MOG BH  in MOG 
theory~\cite{mf,mfjcap,mf1}. It is described by mass parameter~(${\cal M}$), spin parameter~($a$) and 
a deformation parameter or MOG parameter~($\alpha$). In contrast to Kerr BH, which  is described only 
by the mass parameter and the spin parameter. The MOG theory was proposed by Moffat~\cite{mfjcap}. 
It was sometimes reffered to as scalar-tensor-vector gravity (STVG) theory. 
The main concept in MOG theory is that the BH charge parameter is proportional to the mass parameter 
i. e. ${Q}=\sqrt{\alpha G_{N}} M$~\cite{mf}. Where  $\alpha=\frac{G-G_{N}}{G_{N}}$ should 
be measured deviation of MOG from GR. The MOG theory proves several interesting 
features like superradiance~\cite{mf18}, the quasinormal modes and the ring down of BH 
mergers~\cite{mf18plb}. Apart from that MOG theory obeys an action principle formulation, 
and the weak equivalence principle. The details of MOG theory and action formulation could be 
found in the following works~\cite{mf,mf1,mf18,mf18plb}. So, we have not discussed it in detail 
here.

The metric for static spherically-symmetric BH in MOG theory can be obtained by substituting 
${\cal Q}=\sqrt{\alpha G_{N}}M$ in the usual Reissner-Nordstr\"{o}m BH solution. Similarly, 
one can derive the Kerr-MOG BH by putting this condition in  Kerr-Newman BH.

In~\cite{pp20}, we have derived in detail the generalized spin precession frequency of a test 
gyroscope around the Kerr-MOG BH. Using this frequency expression, we differentiated the behavior 
of three compact objects i.e. NXBH, XBH, and NS. Here, we focus on particularly LT frequency by 
taking the limit i.e. angular velocity $\Omega=0$ in generalized spin frequency expression. 

For Kerr BH~\cite{ch17a}(See also \cite{ch17,chur1}), it was demonstrated that a clear 
distinction between these compact objects. In fact they have found that the LT frequency varies as  
$\Omega_{LT}\propto a$  and $\Omega_{LT}\propto \frac{1}{r^3}$.  While for 
Kerr-NS~(or superspinar) and also in the limit $a>>r$~\footnote{The region  $a>>r$ is a typical nature 
of Kerr naked singularities\cite{ch17}.} and in the region $\theta\neq\frac{\pi}{2}$ the 
LT frequency varies as $\Omega_{LT}\propto r$ and $\Omega_{LT}\propto \frac{1}{a^3}$. In the 
present work, we would like to examine this scenario in the MOG theory specifically for
Kerr-MOG BH.

In the present work, we derive the following results:
(i) We show that  for \emph{NXBH} the LT precession frequency along the pole is directly proportional 
to the spin parameter~($a$),  and is inversely proportional to the cubic value of radial distance 
parameter, and 
\begin{equation}
\Omega_{LT} \propto \left[1-\left(\frac{\alpha}{1+\alpha}\right)\frac{G_{N}{\cal M}}{2r}\right]
\left(1-2\frac{G_{N}{\cal M}}{r}+\frac{a^2+\left(\frac{\alpha}{1+\alpha}\right)G_{N}^2{\cal M}^2}{r^2}\right)^{-1} 
~\label{1.1}
\end{equation}

(ii) Along the equatorial plane,  the LT precession frequency  is directly proportional 
to the spin parameter,  and is inversely proportional to the cubic value of radial distance,  and 
\begin{equation}
\Omega_{LT} \propto \left[1-\left(\frac{\alpha}{1+\alpha}\right)\frac{G_{N}{\cal M}}{r}\right]
\left(1-2\frac{G_{N}{\cal M}}{r}+\frac{\left(\frac{\alpha}{1+\alpha}\right)G_{N}^2{\cal M}^2}{r^2}\right)^{-1} 
~\label{1.2}
\end{equation}

(iii) For \emph{XBH}, we determine the LT frequency is proportional to the angular-momentum~($J$) 
parameter i.e. $\Omega_{LT}\propto \frac{{\cal M}^2}{\sqrt{1+\alpha}}$ and is inversely 
proportional to the cubic value of radial parameter i.e. $\Omega_{LT}\propto \frac{1}{r^3}$. 

(iv) While for \emph{superspinar}, the $\Omega_{LT}$  along the pole and in the 
region  $a>>r$ should be
\begin{equation}
\Omega_{LT}\propto \left[r-\left(\frac{\alpha}{1+\alpha}\right)\frac{G_{N}{\cal M}}{2} \right]
\left[1+\left(\frac{\alpha}{1+\alpha}\right)\frac{G_{N}^2{\cal M}^2}{a^2}\right]^{-\frac{1}{2}}
~\label{1.3}
\end{equation}

(v) For $\theta\neq\frac{\pi}{2}$ and in the superspinar limit, we find 
\begin{equation}
\Omega_{LT}\propto \frac{1}{a^3\cos^4\theta}
\end{equation}
and Eq.~(\ref{ns1.2}). Interestingly, the LT precession frequency  is inversely proportional 
to the both $a^3$ and $\cos^4\theta$ factor in the superspinar limit.

There are several discussions regarding the Kerr NS such as stability issue, thermodynamic propeties, 
effect of gravitional self-force~(GSF), effect of conservative self-force and the  implications of 
Polish doughnut model etc. They could be found in the following  
Refs.~\cite{jacobson09,khanna10,joshi14,barak15,saa11,bambi13,defelice78,calvini78,zhang16}.
So, the NS is an interesting topic of research in recent times both from the theoretical 
and observational point of view.

In the next section, we will study the spin precession frequency in Kerr BH in MOG theory and 
show that the structural variation of three compact objects: NXBH, XBH and NS. In Sec.~3, we 
have given the conclusions.

\section{Spin Precession Frequency in Kerr-MOG BH}
To distinguish NXBH, XBH and NS in Kerr-MOG BH, we have to write the LT spin precession 
expression in terms of Boyer-Lindquist coordinates\cite{pp20}[See also \cite{chp,chppp,chppp1,ckd,ch18}] 
as
\begin{eqnarray}
\vec{\Omega}_{LT} &=&  \frac{-\sqrt{g_{rr}}~{\cal A} \hat{r}+\sqrt{g_{\theta\theta}}~{\cal F} 
\hat{\theta}}{2 \sqrt{-g}~g_{tt}}~~\label{n2}
\end{eqnarray}
where 
\begin{eqnarray}
{\cal A} &=& g_{tt}~g_{t\phi,\theta}-g_{t\phi}~g_{tt,\theta}\\
{\cal F} &=& g_{tt}~g_{t\phi,r}-g_{t\phi}~g_{tt,r}
\end{eqnarray}
and $g$ is the determinant of the metric tensor.
This is the exact LT precession frequency of a test gyro due to rotation of any 
stationary and axisymmetric spacetime. 

The most interesting feature of  Eq.~(\ref{n2}) is that it should be applicable 
for any stationary axisymmetric BH spacetime which is valid for both outside and 
inside the ergosphere. Here, we are particularly interested in the limit $\Omega=0$. 
The case of $\Omega\neq 0$ is discussed and analyzed in the work~\cite{pp20}.

Our next task is to write the  metric explicitly for KMOG BH as described in Ref.~\cite{mf}
\begin{eqnarray}
ds^2 = -\frac{\Delta}{\rho^2} \, \left[dt-a\sin^2\theta d\phi \right]^2+\frac{\sin^2\theta}{\rho^2} \,
\left[(r^2+a^2) \,d\phi-a dt\right]^2
+\rho^2 \, \left[\frac{dr^2}{\Delta}+d\theta^2\right] \nonumber\\
~.\label{mg2.1}
\end{eqnarray} 
where
\begin{eqnarray}
\rho^2 & \equiv & r^2+a^2\cos^2~\theta \nonumber\\
\Delta & \equiv & r^2-2G_{N}(1+\alpha)Mr+a^2 + G_{N}^2 \alpha(1+\alpha) M^2 ~.\label{m2.1}
\end{eqnarray}
where $G_{N}$ is Newton's gravitational constant and $M$ is  BH mass. 
We should {mention} that in the metric $c=1$~\footnote{We have used the geometrical units 
i. e. $c=1$. The signature of the metric becomes (-,+,+,+). }. The above metric is an 
axially-symmetric and stationary spacetime.  The ADM mass and angular momentum are computed 
in~\cite{ps}  as ${\cal M}=(1+\alpha)M$ and $J=aG_{N}{\cal M}$ ~\footnote{One could determine 
the relation between the BH mass and ADM mass as $M=\frac{\cal M}{1+\alpha}$.}.
We consider here the ADM mass throughout the paper for our convenience. 
Substituting these values in Eq.~(\ref{m2.1}) then $\Delta$ becomes
\begin{eqnarray}
\Delta & = & r^2-2G_{N}{\cal M}r+a^2 + \frac{\alpha}{1+\alpha} G_{N}^2 {\cal M}^2 
\end{eqnarray}
The above metric describes a BH with double horizon
\begin{eqnarray}
r_{+} &=& G_{N}{\cal M} + \sqrt{\frac{G_{N}^2{\cal M}^2}{1+\alpha}-a^2},\,\, 
r_{-} = G_{N}{\cal M} - \sqrt{\frac{G_{N}^2{\cal M}^2}{1+\alpha}-a^2}~. \label{mg2.2}
\end{eqnarray} 
where $r_{+}$ is called as event horizon and $r_{-}$ is called as Cauchy horizon.

\begin{figure}[h]
\begin{center}
{\includegraphics[width=0.45\textwidth]{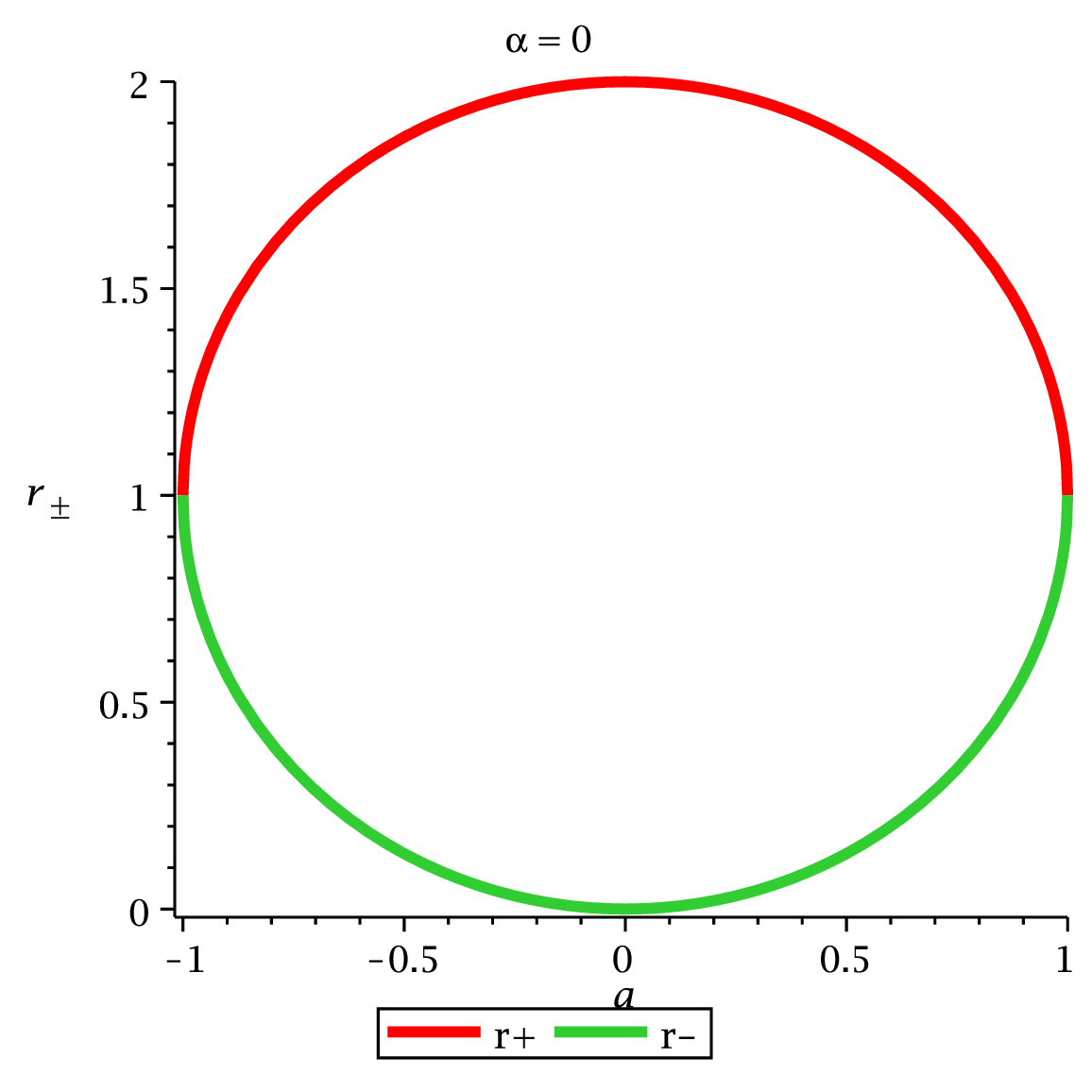}}
{\includegraphics[width=0.45\textwidth]{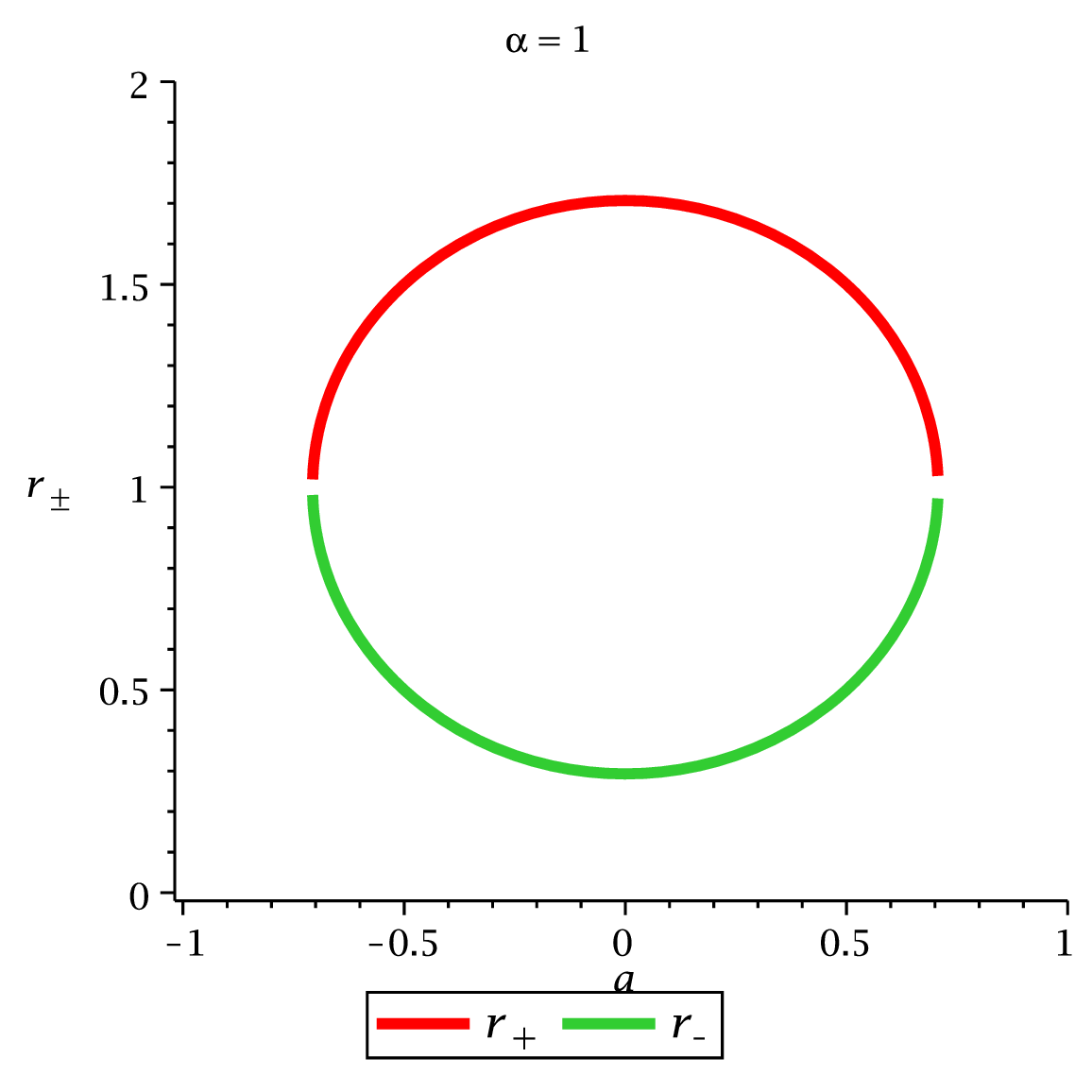}}
\end{center}
\caption{The figure describes the variation  of $r_{\pm}$  with $a$ and $\alpha$ 
for Kerr BH and Kerr-MOG BH. Left figure for Kerr BH. Right figure for Kerr-MOG BH. 
The presence of the MOG parameter is deformed the shape of the horizon radii. 
\label{hr}}
\end{figure}

Note that $r_{+}>r_{-}$. Differences between the horizon structure in the presence of MOG 
parameter and without MOG parameter could be seen visually in the Fig.~\ref{hr}. 
It should be noted that when $\alpha=0$, one gets the horizon radii of Kerr BH.  
The NXBH, XBH and NS solutions do exist in the following range
\begin{eqnarray}
\frac{a^2}{G_{N}^2{\cal M}^2} &<& \frac{1}{{1+\alpha}}~~~~~~{\mbox{NXBH}}\\
\frac{a^2}{G_{N}^2{\cal M}^2} &=& \frac{1}{{1+\alpha}}~~~~~~{\mbox{XBH}}\\
\frac{a^2}{G_{N}^2{\cal M}^2} &>& \frac{1}{{1+\alpha}}~~~~~~{\mbox{NS}}
\end{eqnarray}
Note that the MOG parameter or deformation parameter~($\alpha$) is always positive definite. If we 
invert the above inequality then one gets the restriction on $\alpha$. We have tabulated the values 
of spin parameter for various values of MOG parameter.  
\begin{center}
\begin{tabular}{|c|c|c|c|}
    \hline
    $\alpha$ & NXBH  & XBH & NS\\
    \hline
    $\alpha=0$ & $a=0.5$ & $a=1$ & $a=2$\\
    $\alpha=1$ & $a=0.4$ & $a=\frac{1}{\sqrt{2}}=0.7$ & $a=0.9$\\
    $\alpha=2$ & $a=0.3$ & $a=\frac{1}{\sqrt{3}}=0.57$ & $a=0.7$\\ 
    $\alpha=3$ & $a=0.2$ & $a=\frac{1}{2}=0.5$ & $a=0.8$\\
    $\alpha=4$ & $a=0.2$ & $a=\frac{1}{\sqrt{5}}=0.44$ & $a=0.6$\\ 
    $\alpha=5$ & $a=0.2$ & $a=\frac{1}{\sqrt{6}}=0.4$ & $a=0.5$\\
    $\alpha=6$ & $a=0.2$ & $a=\frac{1}{\sqrt{7}}=0.37$ & $a=0.4$\\
    $\alpha=8$ & $a=0.2$ & $a=\frac{1}{3}=0.33$ & $a=0.5$\\ 
    $\alpha=10$& $a=0.2$ & $a=\frac{1}{\sqrt{11}}=0.3$ & $a=0.5$\\ 
   \hline
\end{tabular}
\end{center}
The outer ergosphere is situated  at 
\begin{eqnarray}
r &=& r_{e}^{+}(\theta) = G_{N}{\cal M} + \sqrt{\frac{G_{N}^2{\cal M}^2}{1+\alpha}-a^2 \cos^2~\theta} 
~. \label{mg2.3}
\end{eqnarray}
and the inner ergosphere is situated  at 
\begin{eqnarray}
r &=& r_{e}^{-}(\theta) = G_{N}{\cal M} - \sqrt{\frac{G_{N}^2{\cal M}^2}{1+\alpha}-a^2 \cos^2~\theta} 
~. \label{mg2.3a}
\end{eqnarray}
and they should satisfy the following inequality $r_{e}^{-}(\theta)\leq r_{-}\leq r_{+}\leq r_{e}^{+}(\theta)$. 
It may be noted that $r_{e}^{+}$ and $r_{e}^{-}$ coincide with $r_{+}$ and $r_{-}$ at the pole. 
The radii of ergo-region becomes imaginary for all values of $\theta$: i.e. 
$\theta>\cos^{-1}\left[\frac{1}{\sqrt{1+\alpha}}\left(\frac{1}{ a_{\ast}}\right)\right]$. 
In the  range $\theta<\cos^{-1}\left[\frac{1}{\sqrt{1+\alpha}}\left(\frac{1}{ a_{\ast}}\right)\right]$ 
the radii of ergo-region becomes real and where $a_{\ast}=\frac{a}{G_{N}{\cal M}}$.

\begin{figure}
\begin{center}
{\includegraphics[width=0.38\textwidth]{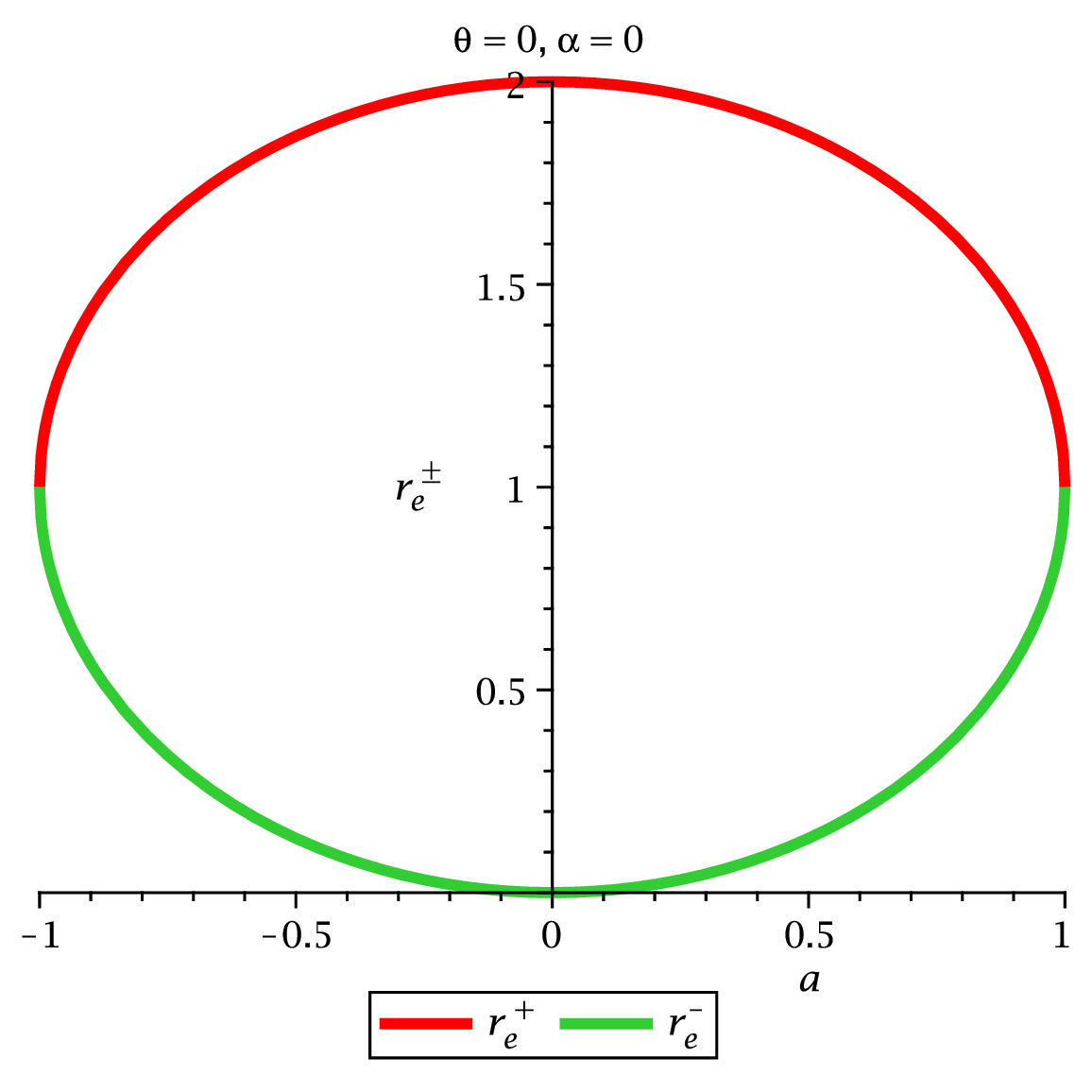}}
{\includegraphics[width=0.38\textwidth]{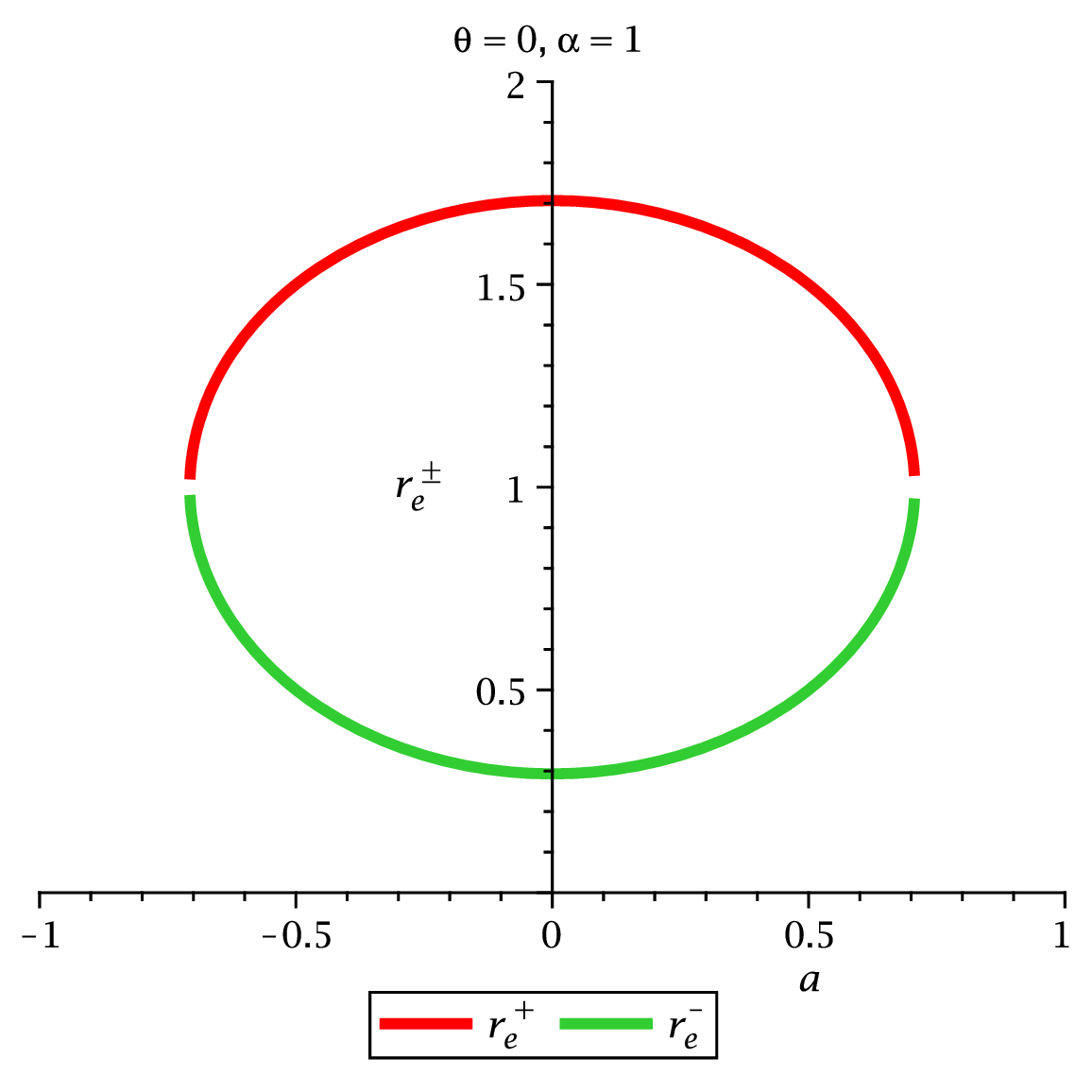}}
{\includegraphics[width=0.38\textwidth]{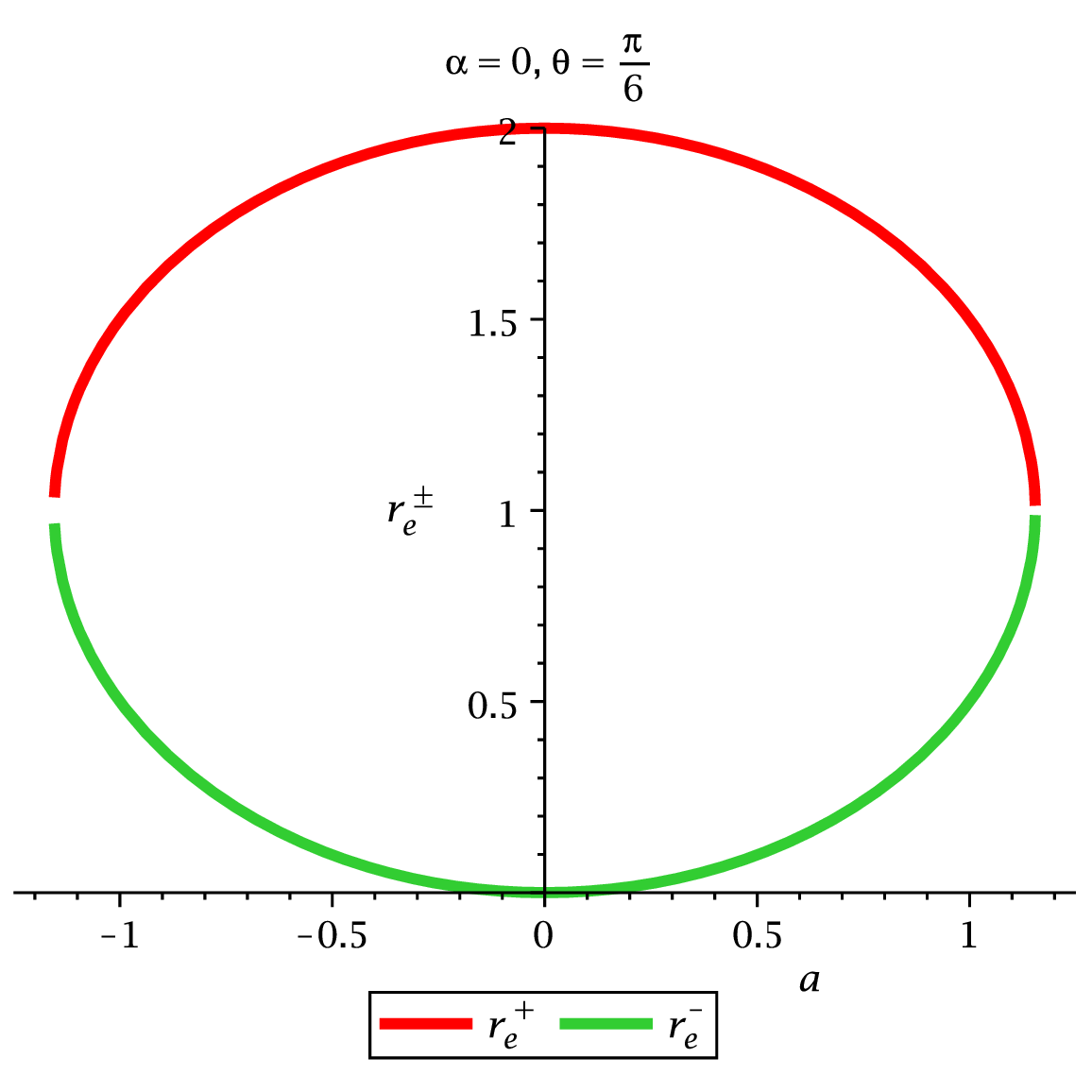}}
{\includegraphics[width=0.38\textwidth]{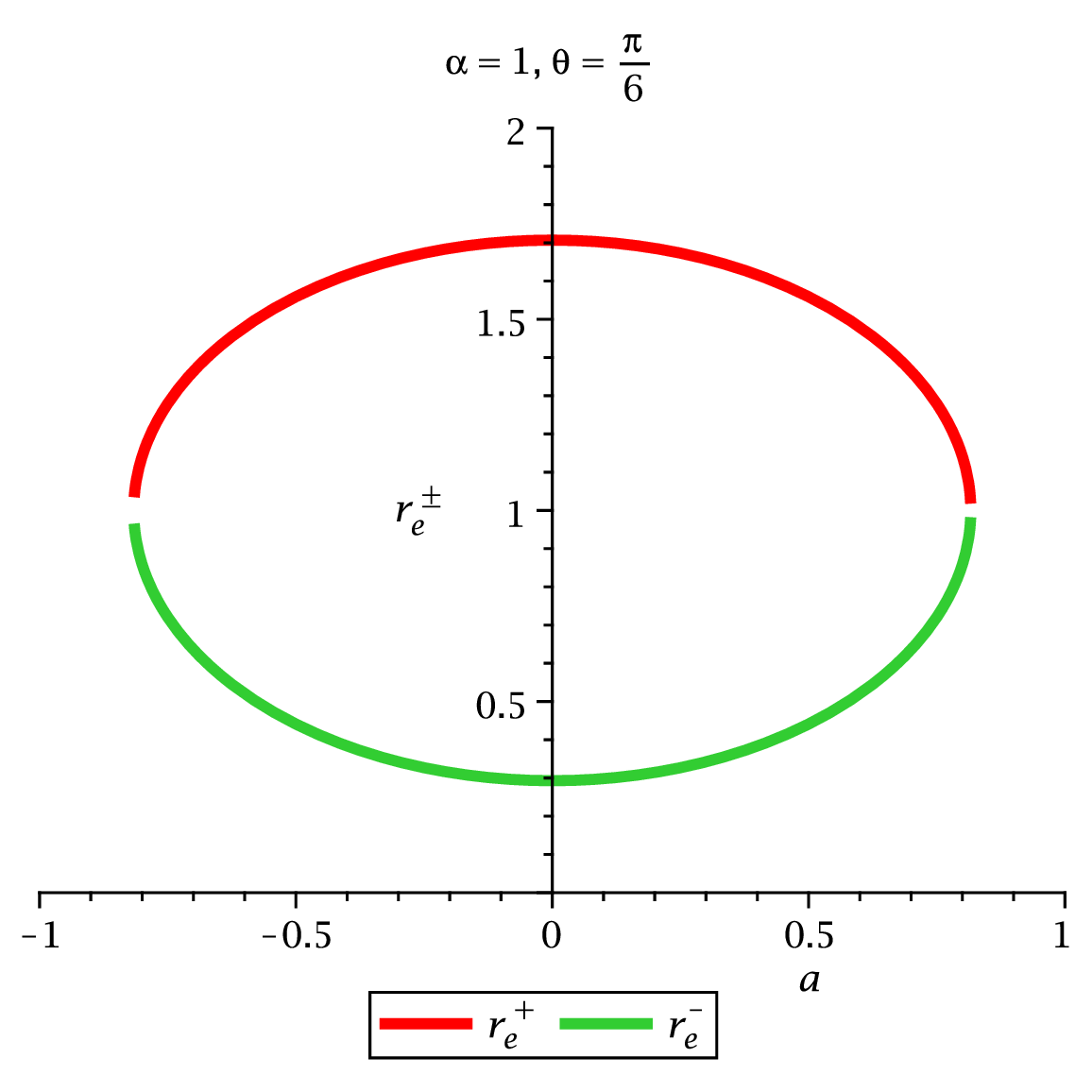}}
\end{center}
\caption{The figure indicates the variation  of $r_{e}^{\pm}(\theta)$ with $a$ and $\alpha$ 
for Kerr BH and Kerr-MOG BH.
\label{erg}}
\end{figure}

\begin{figure}
\begin{center}
{\includegraphics[width=0.38\textwidth]{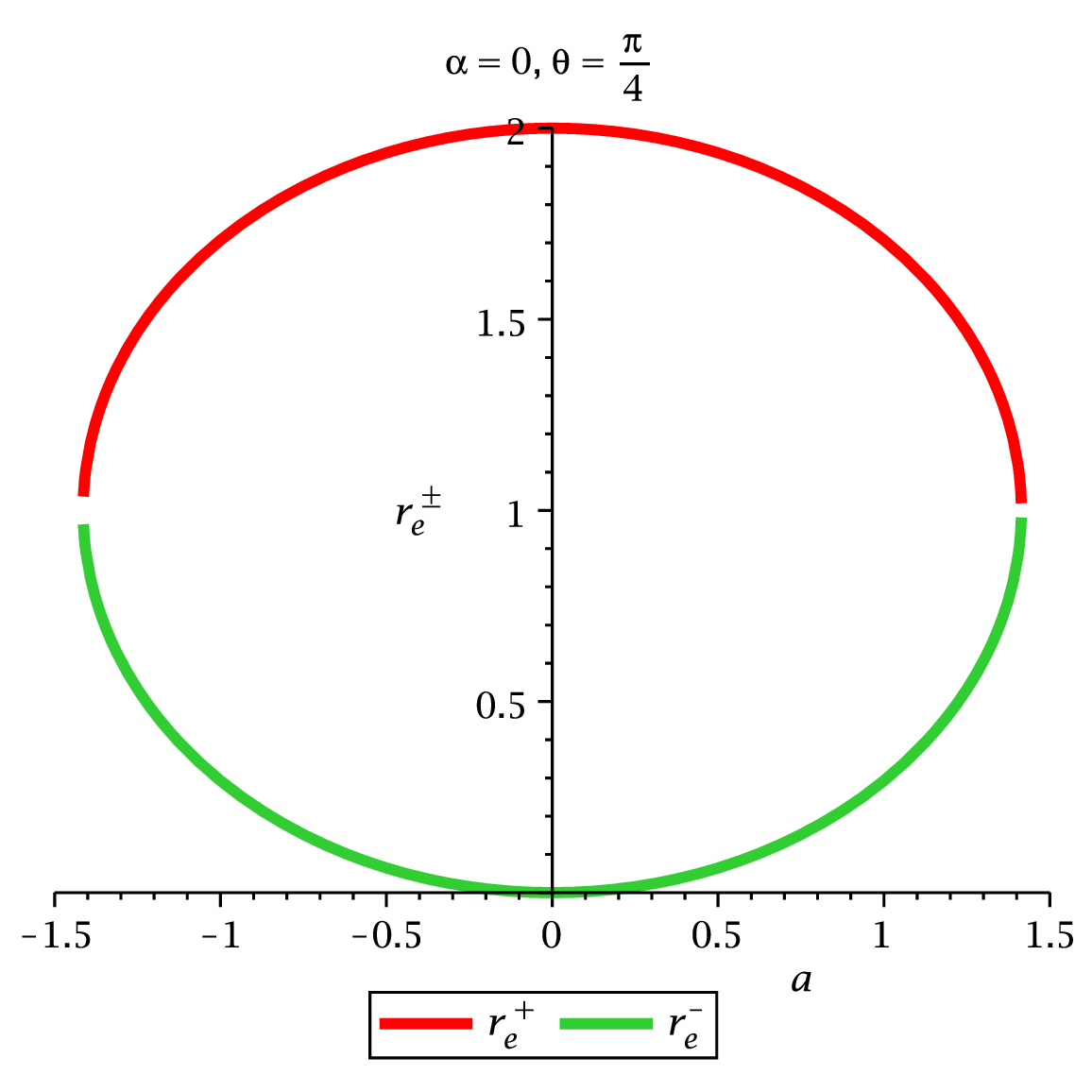}}
{\includegraphics[width=0.38\textwidth]{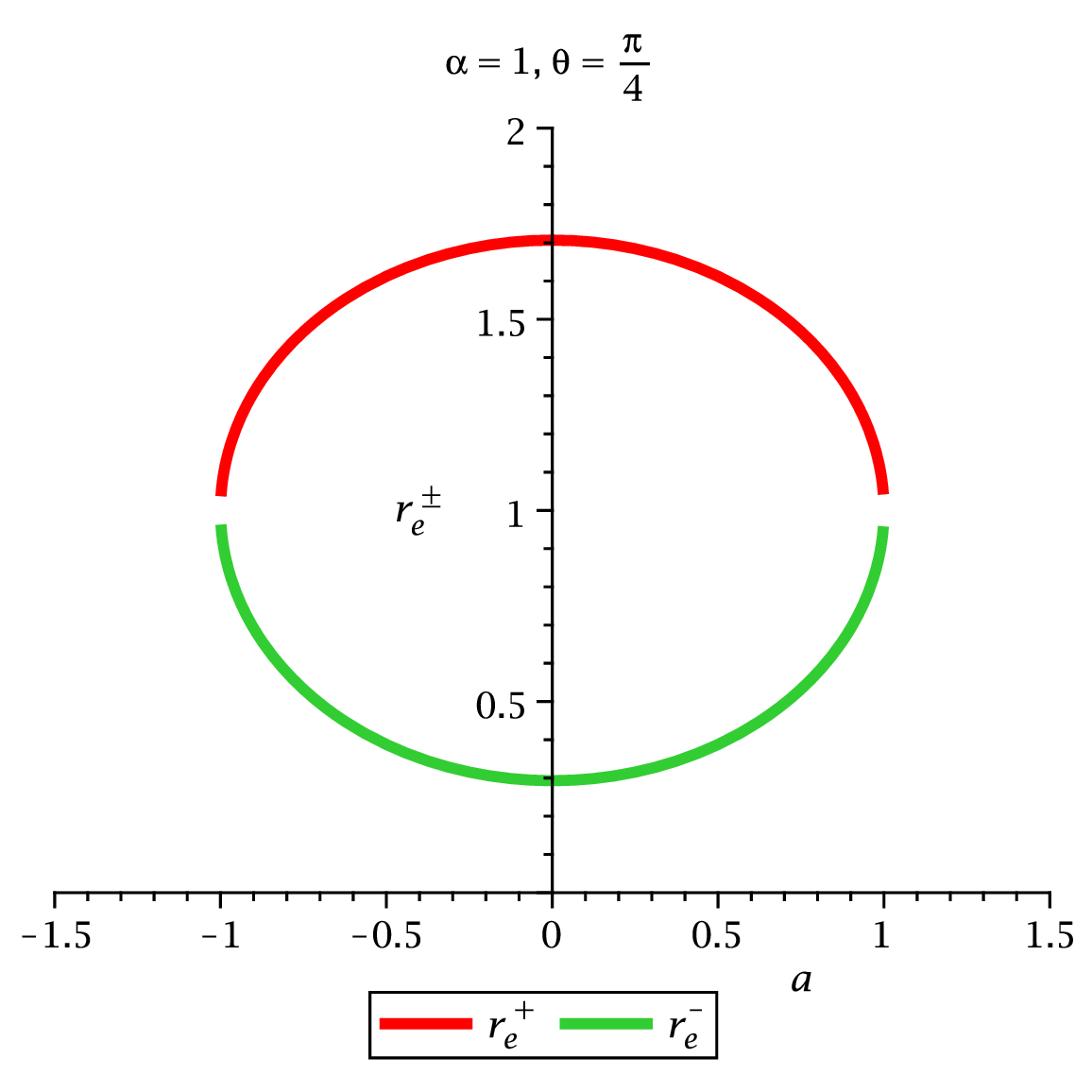}}
{\includegraphics[width=0.38\textwidth]{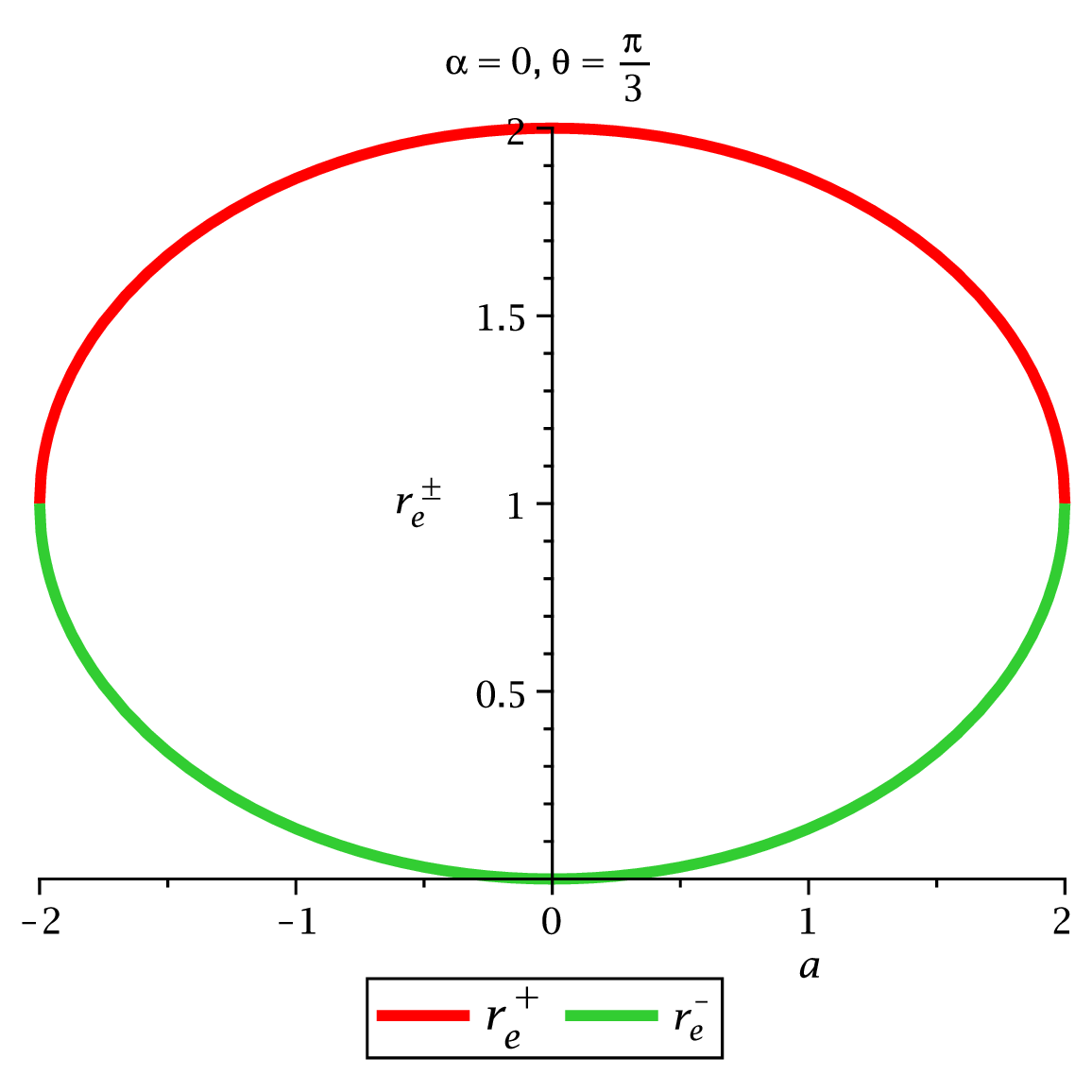}}
{\includegraphics[width=0.38\textwidth]{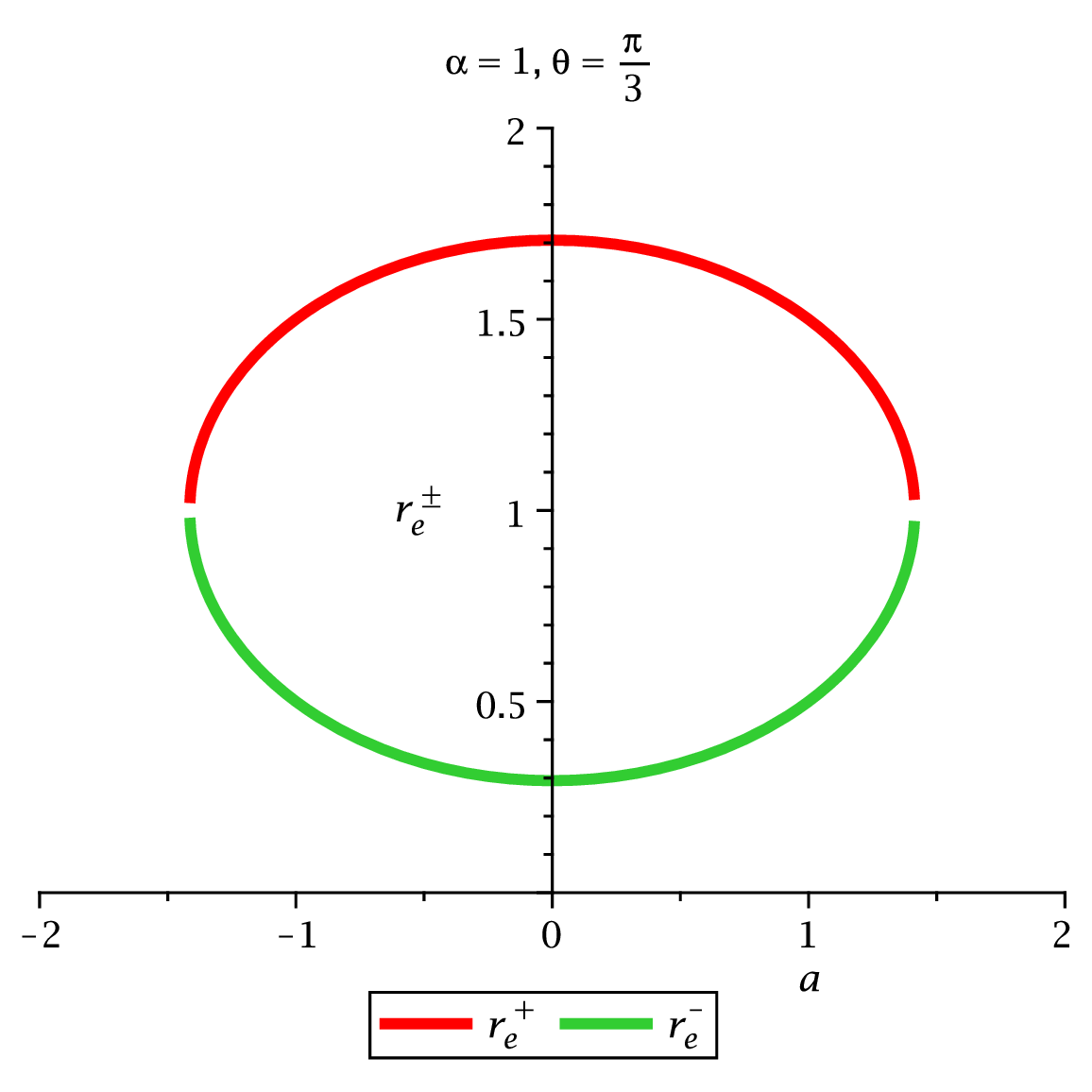}}
\end{center}
\caption{The figure indicates the variation  of $r_{e}^{\pm}(\theta)$ with $a$ and $\alpha$ 
for Kerr BH and Kerr-MOG BH.
\label{erg1}}
\end{figure}

The structural differences of the outer and inner ergosphere in the presence of MOG parameter and 
without MOG parameter could be seen from  Fig.~\ref{erg} and Fig.~\ref{erg1}. From these figures one 
can easily observed that the size of the ergo-sphere increases when one goes to the equator starting 
from pole. It becomes maximum at the equator. 

In the extremal limit, the outer horizon and inner horizon coincide 
at $r_{+}=r_{-}= G_{N}{\cal M}$. Thus the outer and inner ergosphere 
radius reduces to
\begin{eqnarray}
r_{e}^{+}(\theta) &=& G_{N}{\cal M} \left(1 + \frac{\sin\theta}{\sqrt{1+\alpha}}\right),\,\,
r_{e}^{-}(\theta) = G_{N}{\cal M} \left(1 - \frac{\sin\theta}{\sqrt{1+\alpha}}\right)
~. \label{mg2.4}
\end{eqnarray} 
On axis and on equatorial plane, these values are 
\begin{eqnarray}
r_{e}^{\pm}(\theta)|_{\theta=0} &=& G_{N}{\cal M}=r_{\pm}, \,\,\, \mbox{(on axis)}\\
r_{e}^{\pm}(\theta)|_{\theta=\frac{\pi}{2}} &=& G_{N}{\cal M} \left(1 \pm \frac{1}{\sqrt{1+\alpha}}\right)
=r_{\pm}|_{a=0}\,\,\,\mbox{(equatorial plane)}~. \label{mg2.5}
\end{eqnarray}
In the limit $\alpha=0$, 
\begin{eqnarray}
r_{e}^{+}(\theta)|_{\theta=\frac{\pi}{2}} &=& 2G_{N}{\cal M}, \,\,\,\, r_{e}^{-}(\theta)|_{\theta=\frac{\pi}{2}}=0 
~. \label{mg2.6}
\end{eqnarray}
This surface is outer to the event horizon or outer horizon and it coincides with the outer horizon at 
the poles $\theta=0$ and $\theta=\pi$. 

Now using the formula~(\ref{n2}), one obtains the LT frequency vector for the metric~(\ref{mg2.1})
\begin{eqnarray}
\vec{\Omega}_{LT} &=& \frac{\chi(r)~\sqrt{\Delta}\cos\theta~\hat{r}+\mu(r)~\sin\theta~\hat{\theta}}{\sigma(r)}
,~\label{gekm3}
\end{eqnarray}
where
$$
\chi(r)= a \Pi_{\alpha}
$$
$$
\Pi_{\alpha}=2G_{N}{\cal M} r-\frac{\alpha}{1+\alpha} G_{N}^2 {\cal M}^2
$$

$$
\mu(r) = aG_{N}{\cal M}(r^2-a^2\cos^2\theta)-\frac{\alpha}{1+\alpha} G_{N}^2 {\cal M}^2 ar
$$
$$
\sigma(r) = \rho^3(\rho^2-\Pi_{\alpha})
$$
The magnitude of this vector is computed to be 
\begin{eqnarray}
\Omega_{LT}(r,\theta) &=& \frac{\sqrt{\Delta~\chi^2(r)~\cos^2\theta+\mu^2(r)~\sin^2\theta}}
{\sigma(r)} ~\label{ttl3}
\end{eqnarray}
After substituting the values of $\Delta$, $\mu(r)$, $\chi(r)$, $\sigma(r)$ and $\rho$, we get 
the LT precession frequency for \emph{NXBH} is 
\begin{equation}
\Omega_{LT}= \frac{J}{r^3}\,\Upsilon(r,a,\theta,\alpha)
\end{equation}
where 
$$
\Upsilon(r,a,\theta,\alpha)\equiv \Upsilon
$$
and 
\begin{eqnarray}
\Upsilon=\frac{\sqrt{4\left[1-\left(\frac{\alpha}{1+\alpha}\right)\frac{G_{N}{\cal M}}{2r}\right]^2
\left[1-2\frac{G_{N}{\cal M}}{r}+\frac{a^2+\left(\frac{\alpha}{1+\alpha}\right)G_{N}^2{\cal M}^2}{r^2}\right]\cos^2\theta 
+\sin^2\theta \left[1-\frac{a^2\cos^2\theta}{r^2}-\left(\frac{\alpha}{1+\alpha}\right)\frac{G_{N}{\cal M}}{r}\right]^2}}
{\left(1+\frac{a^2\cos^2\theta}{r^2}\right)^\frac{3}{2} 
\left(1-2\frac{G_{N}{\cal M}}{r}+\frac{a^2\cos^2\theta+\left(\frac{\alpha}{1+\alpha}\right)G_{N}^2{\cal M}^2}{r^2}\right)}
\nonumber\\  
\end{eqnarray}

Now we will compute the precession frequency along the pole and equatorial plane  for NXBH separately.\\
Case I: 
At the pole~($\theta=0$), the LT precession frequency is 
\begin{equation}
\Omega_{LT}= \frac{2J}{r^3}\left[1-\left(\frac{\alpha}{1+\alpha}\right)\frac{G_{N}{\cal M}}{2r}\right]
\left(1+\frac{a^2}{r^2}\right)^{-\frac{3}{2}} 
\left(1-2\frac{G_{N}{\cal M}}{r}+\frac{a^2+\left(\frac{\alpha}{1+\alpha}\right)G_{N}^2{\cal M}^2}{r^2}\right)^{-1}  
\end{equation}
It indicates that the LT precession frequency along the pole is directly proportional  to the angular-momentum 
parameter,  and is inversely proportional to the cubic value of radial distance and 
\begin{equation}
\Omega_{LT} \propto \left[1-\left(\frac{\alpha}{1+\alpha}\right)\frac{G_{N}{\cal M}}{2r}\right]
\left(1-2\frac{G_{N}{\cal M}}{r}+\frac{a^2+\left(\frac{\alpha}{1+\alpha}\right)G_{N}^2{\cal M}^2}{r^2}\right)^{-1}  
\end{equation}

Case II: 
At the equatorial plane~($\theta=\frac{\pi}{2}$), the LT precession frequency is 
\begin{equation}
\Omega_{LT}= \frac{J}{r^3}\left[1-\left(\frac{\alpha}{1+\alpha}\right)\frac{G_{N}{\cal M}}{r}\right]
\left(1-2\frac{G_{N}{\cal M}}{r}+\frac{\left(\frac{\alpha}{1+\alpha}\right)G_{N}^2{\cal M}^2}{r^2}\right)^{-1} 
\end{equation}
Analogously, at the equatorial plane the LT precession frequency  is directly proportional 
to the angular-momentum parameter,  and is inversely proportional to the cubic value of radial distance 
and 
\begin{equation}
\Omega_{LT} \propto \left[1-\left(\frac{\alpha}{1+\alpha}\right)\frac{G_{N}{\cal M}}{r}\right]
\left(1-2\frac{G_{N}{\cal M}}{r}+\frac{\left(\frac{\alpha}{1+\alpha}\right)G_{N}^2{\cal M}^2}{r^2}\right)^{-1} 
\end{equation}

For \emph{XBH}, the LT frequency is derived to be 
\begin{equation}
\Omega_{LT}= \frac{\Upsilon(r,\theta,\alpha)}{r^3} \frac{G_{N}^2{\cal M}^2}{\sqrt{1+\alpha}}
\end{equation}
where 
\begin{eqnarray}
\Upsilon(r,\theta,\alpha)=\frac{\sqrt{4\left[1-\left(\frac{\alpha}{1+\alpha}\right)\frac{G_{N}{\cal M}}{2r}\right]^2
\left(1-\frac{G_{N}{\cal M}}{r}\right)^2\cos^2\theta +\sin^2\theta \left[1-\frac{G_{N}^2{\cal M}^2}{1+\alpha}
\frac{\cos^2\theta}{r^2}-\left(\frac{\alpha}{1+\alpha}\right)\frac{G_{N}{\cal M}}{r}\right]^2}}
{\left(1+\frac{G_{N}^2{\cal M}^2}{1+\alpha}\frac{\cos^2\theta}{r^2}\right)^\frac{3}{2} 
\left[1-2\frac{G_{N}{\cal M}}{r}+\frac{G_{N}^2{\cal M}^2}{r^2} \left(\frac{\cos^2\theta+\alpha}{1+\alpha} \right)\right]}
\nonumber\\  
\end{eqnarray}
Similarly, we compute the precession frequency along the pole and equatorial plane separately.\\
Case I: 
Along the pole~($\theta=0$), the LT precession frequency is 
\begin{equation}
\Omega_{LT}=\frac{G_{N}^2{\cal M}^2}{\sqrt{1+\alpha}} \left(\frac{2}{r^3}\right)
\left[1-\left(\frac{\alpha}{1+\alpha}\right)\frac{G_{N}{\cal M}}{2r}\right]
\left(1+\frac{G_{N}^2{\cal M}^2}{1+\alpha}\frac{1}{r^2}\right)^{-\frac{3}{2}}
\left(1-\frac{G_{N}{\cal M}}{r}\right)^{-1}  
\end{equation}
Case II: 
Along the equatorial plane~($\theta=\frac{\pi}{2}$), the LT precession frequency is 
\begin{equation}
\Omega_{LT}= \frac{G_{N}^2{\cal M}^2}{\sqrt{1+\alpha}}\left(\frac{1}{r^3}\right)
\left[1-\left(\frac{\alpha}{1+\alpha}\right)\frac{G_{N}{\cal M}}{r}\right]
\left(1-2\frac{G_{N}{\cal M}}{r}+\frac{\left(\frac{\alpha}{1+\alpha}\right)G_{N}^2{\cal M}^2}{r^2}\right)^{-1} 
~\label{ns1.6}
\end{equation}
In each case we observe that the LT precession frequency  is directly proportional 
to the $\frac{{\cal M}^2}{\sqrt{1+\alpha}}$,  and is inversely proportional to the cubic value of 
radial distance parameter.

Finally for \emph{NS} and in the  regime $a>>r$,  the LT frequency is computed to be 
\begin{equation}
\Omega_{LT}= \frac{\sqrt{\left[a^2+\left(\frac{\alpha}{1+\alpha}\right)G_{N}^2{\cal M}^2\right]
\left[2G_{N}{\cal M} r-\left(\frac{\alpha}{1+\alpha}\right)G_{N}^2{\cal M}^2\right]^2\cos^2\theta
+\left[G_{N}\,{\cal M}\,a^2\cos^2\theta+\left(\frac{\alpha}{1+\alpha}\right)G_{N}^2{\cal M}^2\,r \right]^{2} \sin^2\theta}}
{a^2\cos^3\theta\left[a^2\cos^2\theta+\left(\frac{\alpha}{1+\alpha}\right)G_{N}^2\,{\cal M}^2\right]}~\label{ns1.5}
\end{equation} 
and along  the pole 
\begin{equation}
\Omega_{LT}= \frac{2G_{N}{\cal M}\left[r-\left(\frac{\alpha}{1+\alpha}\right)\frac{G_{N}{\cal M}}{2} \right]}
{a^3\sqrt{1+\left(\frac{\alpha}{1+\alpha}\right)\frac{G_{N}^2{\cal M}^2}{a^2}}}
\end{equation}
This immediately implies that 
\begin{equation}
\Omega_{LT} \propto \frac{1}{a^3} 
\end{equation}
and 
\begin{equation}
\Omega_{LT} \propto \left[r-\left(\frac{\alpha}{1+\alpha}\right)\frac{G_{N}{\cal M}}{2} \right]
\left[1+\left(\frac{\alpha}{1+\alpha}\right)\frac{G_{N}^2{\cal M}^2}{a^2}\right]^{-\frac{1}{2}}~\label{ns1.4}
\end{equation}

For $\theta\neq\frac{\pi}{2}$ and in the superspinar limit, we find the exact LT frequency
$$
\Omega_{LT}= \frac{G_{N}{\cal M}\left[r-\left(\frac{\alpha}{1+\alpha}\right)\frac{G_{N}{\cal M}}{2} \right]}{a^3\cos^4\theta}
\left[1+\left(\frac{\alpha}{1+\alpha}\right)\frac{G_{N}^2{\cal M}^2}{a^2\cos^2\theta}\right]^{-1}\times
$$
\begin{equation}
\sqrt{4\left\{1+\left(\frac{\alpha}{1+\alpha}\right)\frac{G_{N}^2{\cal M}^2}{a^2} \right\}+
\left(\frac{\tan\theta}{a}\right)^2\left(\frac{\alpha}{1+\alpha}\right)^2G_{N}^2{\cal M}^2 
\frac{\left\{r+\left(\frac{1+\alpha}{\alpha}\right)\frac{a^2\cos^2\theta}{G_{N}{\cal M}}\right\}^2} 
{\left(r-\left(\frac{\alpha}{1+\alpha}\right)\frac{G_{N}{\cal M}}{2}\right)^{2}}}~\label{ns1.3}
\end{equation}
that means
$$
\Omega_{LT}\propto \left[r-\left(\frac{\alpha}{1+\alpha}\right)\frac{G_{N}{\cal M}}{2} \right]
\left[1+\left(\frac{\alpha}{1+\alpha}\right)\frac{G_{N}^2{\cal M}^2}{a^2\cos^2\theta}\right]^{-1}\times
$$
\begin{equation}
\sqrt{4\left\{1+\left(\frac{\alpha}{1+\alpha}\right)\frac{G_{N}^2{\cal M}^2}{a^2} \right\}+
\left(\frac{\tan\theta}{a}\right)^2\left(\frac{\alpha}{1+\alpha}\right)^2G_{N}^2{\cal M}^2 
\frac{\left\{r+\left(\frac{1+\alpha}{\alpha}\right)\frac{a^2\cos^2\theta}{G_{N}{\cal M}}\right\}^2} 
{\left(r-\left(\frac{\alpha}{1+\alpha}\right)\frac{G_{N}{\cal M}}{2}\right)^{2}}} ~\label{ns1.2}
\end{equation}
and 
\begin{equation}
\Omega_{LT}\propto \frac{1}{a^3\cos^4\theta}~\label{ns1.1}
\end{equation}

\begin{figure}[h]
\begin{center}
\subfigure[]{
\includegraphics[width=1.6in,angle=0]{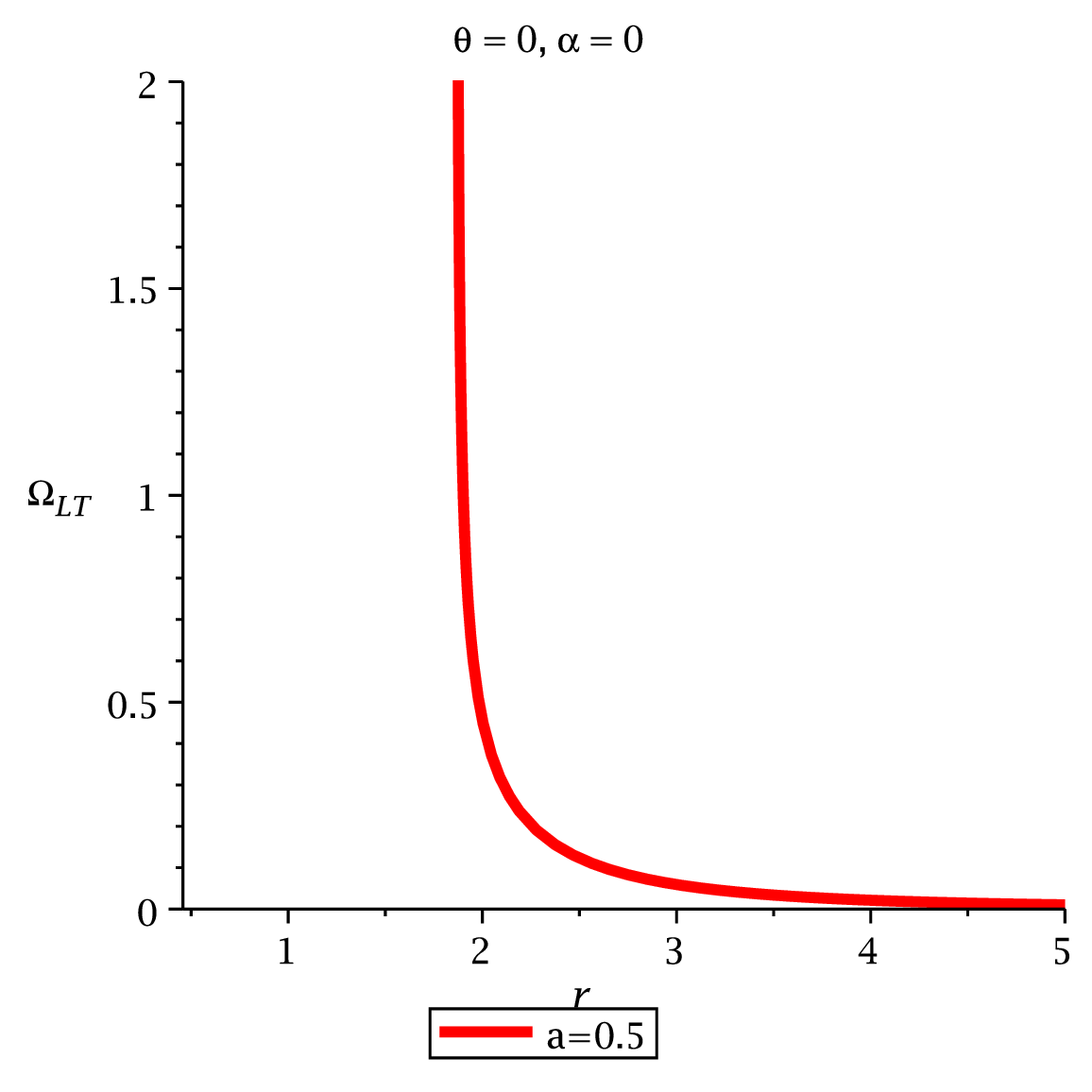}} 
\subfigure[]{
\includegraphics[width=1.6in,angle=0]{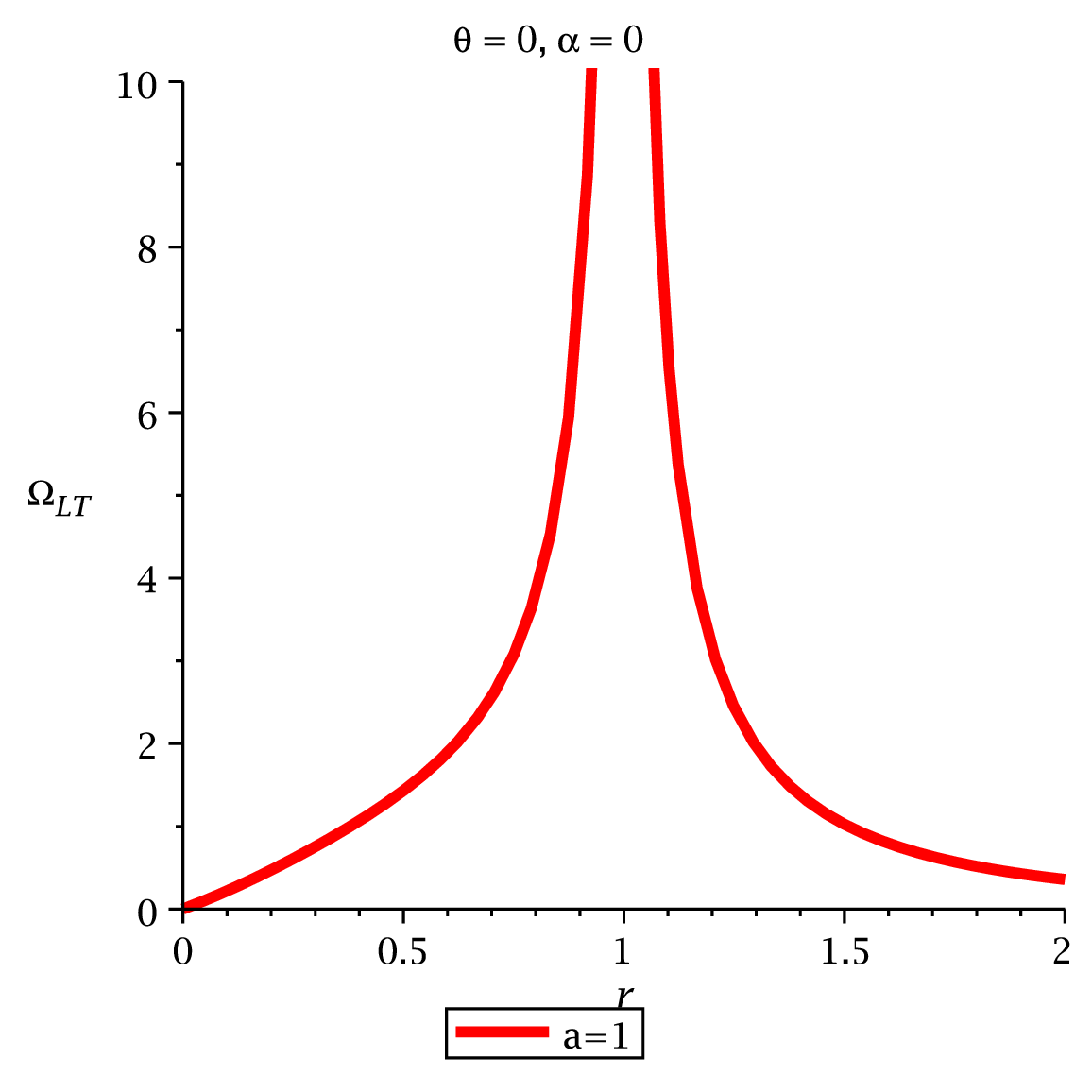}} 
\subfigure[]{
\includegraphics[width=1.6in,angle=0]{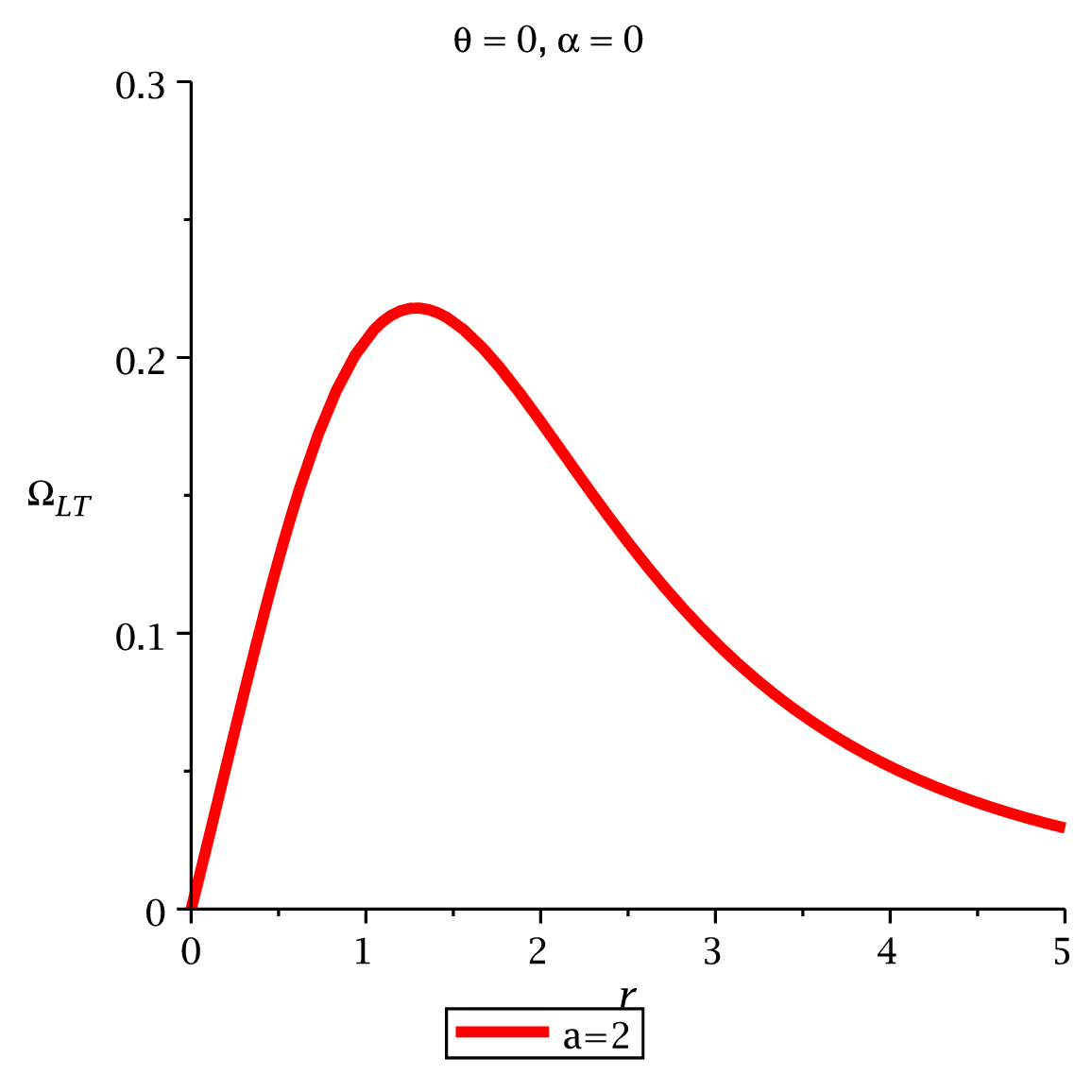}} 
\subfigure[]{
\includegraphics[width=1.6in,angle=0]{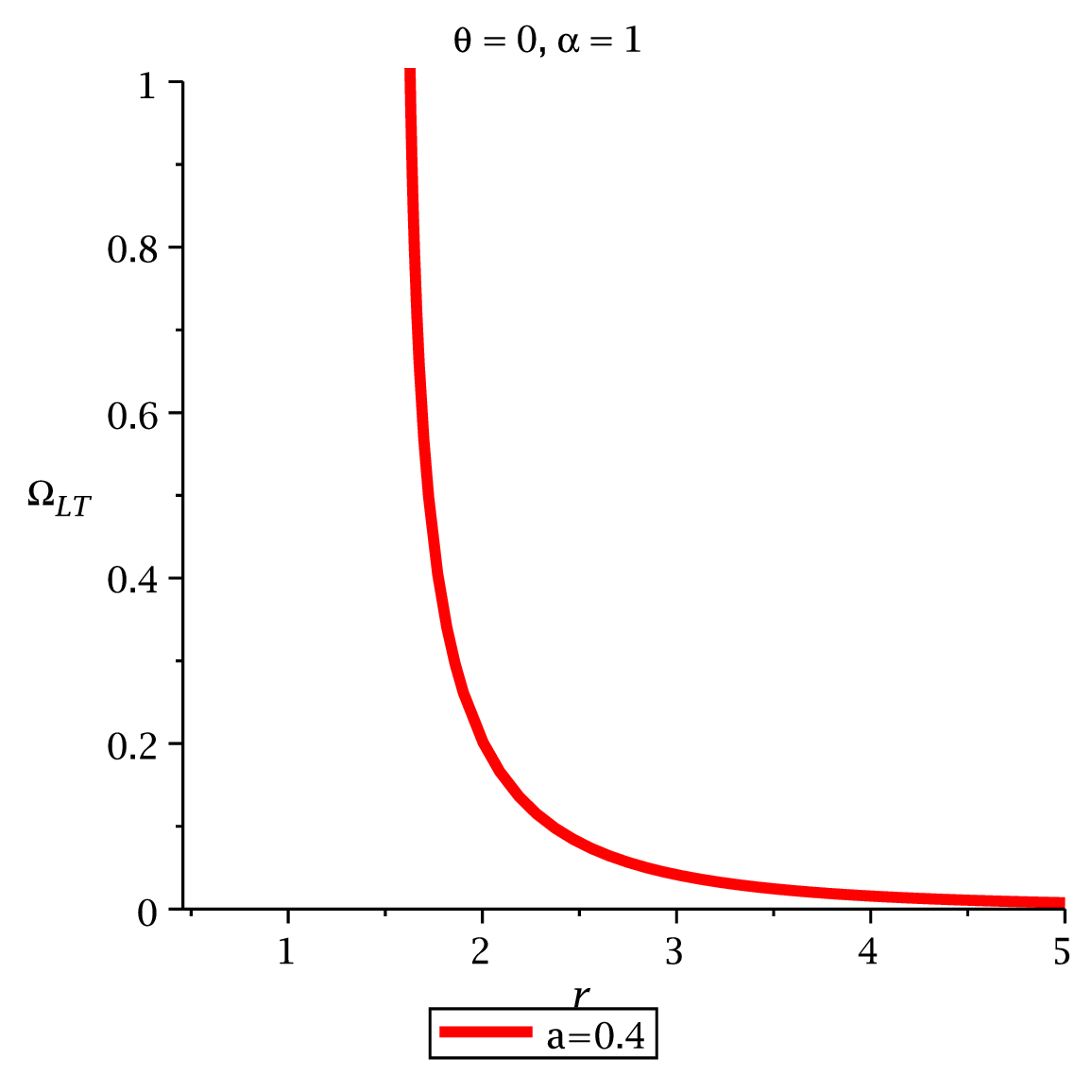}} 
\subfigure[]{
\includegraphics[width=1.6in,angle=0]{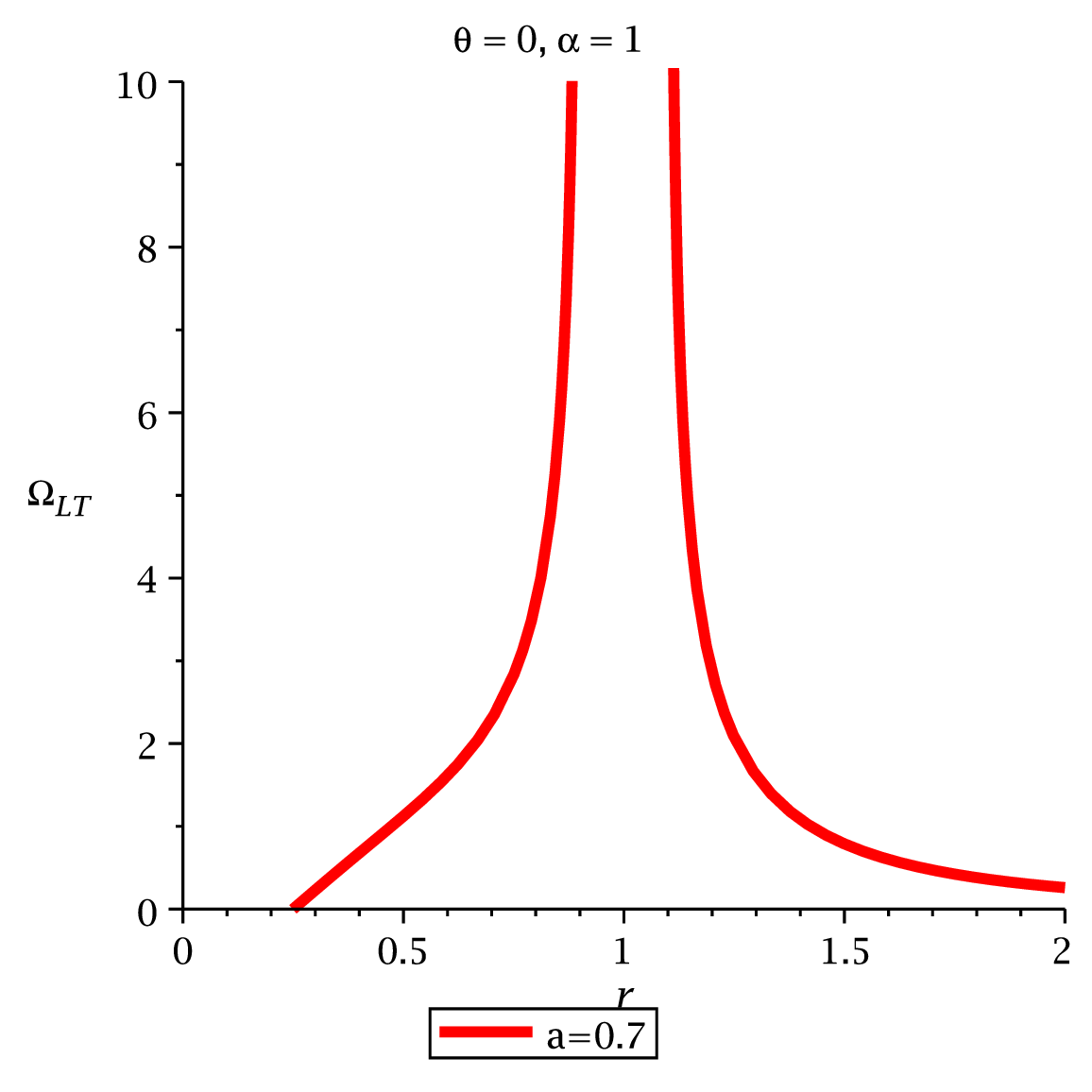}} 
\subfigure[]{
\includegraphics[width=1.6in,angle=0]{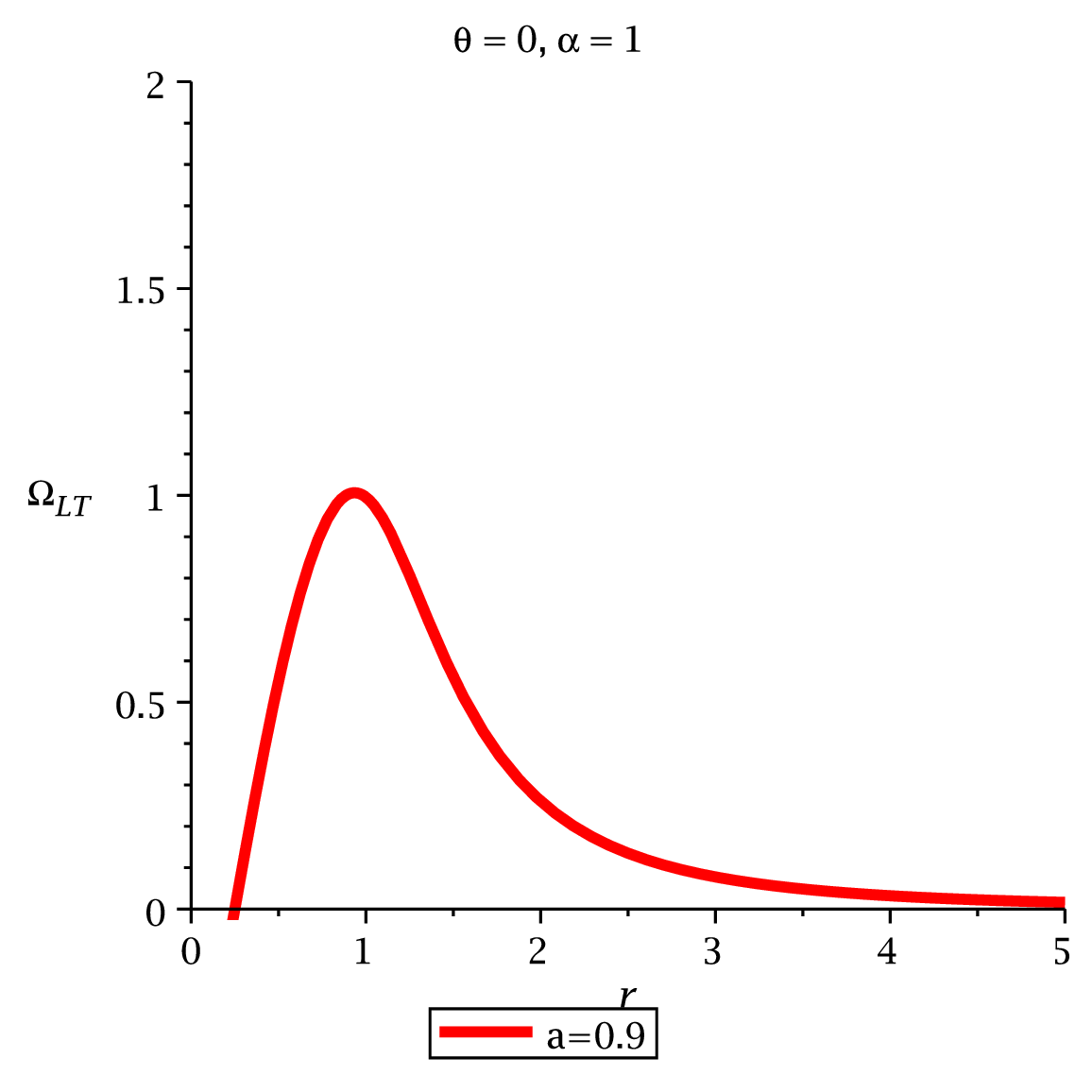}} 
\end{center}
\caption{ The structural variation  of $\Omega_{LT}$ versus $r$ for $\theta=0$ in 
KMOG with variation of MOG parameter and  spin parameter. The first figure describes 
the variation  of $\Omega_{LT}$  with $r$ for NXBH, XBH and NS without MOG parameter. 
The rest of the figure describes the variation  of $\Omega_{LT}$  with $r$ for NXBH, 
XBH and NS with MOG parameter.  Using these plots one can easily distinguish these 
three compact objects} \label{gm}
\end{figure}

These are the fundamental differences between the NXBH, XBH and NS via computation of LT frequency.
Also this is  the main result of our work. Now we have to see their structural differences based on 
the graphical plot~[Fig.~(\ref{gm})]

\section{Discussion}
We studied  the geometrical difference between NXBH and superspinar of Kerr BH in MOG theory 
in terms of precession of the spin of a test gyroscope due to the frame-dragging effect by 
the central rotating body. We found that there is indeed an crucial difference between these 
compact objects in the behavior of gyro spin precession frequency. For NXBH, we have found 
that the LT precession frequency varies as $\Omega_{LT}\propto a $ and  $\Omega_{LT}\propto \frac{1}{r^3}$. 
While for XBH, the LT precession frequency varies as $\Omega_{LT}\propto \frac{{\cal M}^2}{\sqrt{1+\alpha}}$ 
and  $\Omega_{LT}\propto \frac{1}{r^3}$. Note that the behavior of LT gyro precession of NXBH is 
qualitatively same as XBH. For NS, it varies as $\Omega_{LT}\propto \frac{1}{a^3}$
and $\Omega_{LT}\propto \left[r-\left(\frac{\alpha}{1+\alpha}\right)\frac{G_{N}{\cal M}}{2} \right]
\left[1+\left(\frac{\alpha}{1+\alpha}\right)\frac{G_{N}^2{\cal M}^2}{a^2}\right]^{-\frac{1}{2}}$ 
in the regime $\theta=0$ and $a>>r$ limit. For $\theta\neq\frac{\pi}{2}$, we have found Eq.~(\ref{ns1.2}) and 
$\Omega_{LT}\propto \frac{1}{a^3\cos^4\theta}$. Using these specific criterion one can differentiate 
these compact objects. In summary, for a compact object like BH, the LT frequency varies as  
$\Omega_{LT}\propto a $ and  $\Omega_{LT}\propto \frac{1}{r^3}$. While for superspinar, the LT frequency 
varies as $\Omega_{LT}\propto \frac{1}{a^3}$ and 
$\Omega_{LT}\propto \left[r-\left(\frac{\alpha}{1+\alpha}\right)\frac{G_{N}{\cal M}}{2} \right]
\left[1+\left(\frac{\alpha}{1+\alpha}\right)\frac{G_{N}^2{\cal M}^2}{a^2}\right]^{-\frac{1}{2}}$ along 
the pole. In the $\theta\neq\frac{\pi}{2}$ limit, the spin frequency is governed by Eq.~(\ref{ns1.2}) and 
also $\Omega_{LT}\propto \frac{1}{a^3\cos^4\theta}$ .
It is unlikely that Kerr BH in Einstein's general relativity where the precession frequency varies 
as only radial distance parameter.

\end{document}